\documentclass[onecollarge]{svjour2}                      
\smartqed  
\usepackage{graphicx}
\newcommand{\be}{\begin{equation}}
\newcommand{\ee}{\end{equation}}
\newcommand{\bea}{\begin{eqnarray}}
\newcommand{\eea}{\end{eqnarray}}
\newcommand{\f}{\frac}
\newcommand{\s}{\sqrt}

\newcommand{\e}{\epsilon}
\newcommand{\n}{\neq}

\journalname{Gen Relativ Gravit}
\begin{document}

\title{Scalar field exact solutions for non-flat FLRW cosmology:  A technique from non-linear Schr\"{o}dinger-type
formulation \thanks{This work is supported by a TRF-CHE Research
Career Development Grant of the Thailand Research Fund and the
Naresuan Faculty of Science Research Scheme.} }
\titlerunning{Scalar field exact solutions for non-flat FLRW cosmology:  A technique from non-linear Schr\"{o}dinger-type
formulation}        

\author{Burin Gumjudpai}

\authorrunning{Burin Gumjudpai} 

\institute{Burin Gumjudpai \at Fundamental Physics \& Cosmology
Research Unit, The Tah Poe Academia Institute (TPTP)\\ Department of
Physics, Naresuan University, Phitsanulok 65000, Thailand\\ and\\
School of Physics, Institute of Science, Suranaree University of
Technology\\ 111 University Avenue, Nakhon Ratchasima 30000,
Thailand \\
\email{buring@nu.ac.th}}           

\date{Received: date / Accepted: date}

\maketitle

\begin{abstract}

We report a method of solving for canonical scalar field exact
solution in a non-flat FLRW universe with barotropic fluid using
non-linear Schr\"{o}dinger (NLS)-type formulation in comparison to
the method in the standard Friedmann framework. We consider phantom
and non-phantom scalar field cases with exponential and power-law
accelerating expansion. Analysis on effective equation of state to
both cases of expansion is also performed. We speculate and comment
on some advantage and disadvantage of using the NLS formulation in
solving for the exact solution.

 \keywords{Scalar field cosmology \and Non-linear Schr\"{o}dinger
 equation \and Power-law expansion \and Exponential expansion} \PACS{98.80.Cq}

\end{abstract}

\section{Introduction}
\label{sec:introduction} In the past decade, it has been observed
that universe is now in acceleration phase
\cite{Scranton:2003in,Riess:1998cb,Riess:2004nr,Astier:2005qq} while
inflationary scenario of the early universe \cite{inflation} is
strongly confirmed by cosmic microwave background data
\cite{Spergel:2006hy,Hinshaw:2008kr,Dunkley:2008ie,Komatsu:2008hk}.
In both circumstances, the universe experiences accelerating
expansion which can be attained by exploiting some dynamical scalar
field with time-dependent equation of state coefficient $w_{\phi}(t)
< -1/3$, or a cosmological constant with $w_{\Lambda} = -1$
\cite{Padmanabhan:2004av}. It was further suggested that the scalar
field could have phantom behavior with $w_{\phi} <-1$ as one
considers kinetic energy term in its Lagrangian density to be
negative \cite{Caldwell:1999ew,Melchiorri:2002ux}. Strong supports
to the phantom idea are from observations previously, e.g. combined
cosmic microwave background, large scale structure survey and
supernovae type Ia without assuming flat universe yields $
w_{\phi,0} = -1.06^{+0.13}_{-0.08}\;$ \cite{Spergel:2006hy} while
using supernovae data alone assuming flat universe,
$w_{\phi,0}=-1.07\pm 0.09\;$ \cite{WoodVasey:2007jb}. The subscript
0 denotes the value at present. Moreover, most recent WMAP five-year
result \cite{Hinshaw:2008kr,Dunkley:2008ie} combined with Baryon
Acoustic Oscillation (BAO) of large scale structure survey from SDSS
and 2dFGRS \cite{Percival:2007yw} and type Ia supernovae data from
HST \cite{Riess:2004nr}, SNLS \cite{Astier:2005qq} and ESSENCE
\cite{WoodVasey:2007jb} assuming dynamical $w$ with flat universe
yields $-1.38 < w_{\phi,0} < -0.86$ at 95\% confident level and
$w_{\phi,0} = -1.12 \pm 0.13$ at 68\% confident level
\cite{Komatsu:2008hk}. With additional BBN constraint of limit of
expansion rate \cite{Steigman:2007xt,Wright:2007vr}, $-1.32 <
w_{\phi,0} < -0.86$ at 95\% confident level and $w_{\phi,0} = -1.09
\pm 0.12$ at 68\% confident level. This suggests that phantom field
has firmed status in cosmology. However, phantom field does result
in unwanted Big Rip singularity in a flat FLRW  universe
\cite{Caldwell:2003vq}. This raises up many attempts to avoid the
singularity based on both phenomenological and fundamental
inspirations \cite{Sami:2005zc}.

Recently, there are a few proposals for mathematical alternatives to
the conventional Friedmann formulation of canonical scalar field
cosmology, such as non-linear Ermakov-Pinney equation
\cite{Hawkins:2001zx,Williams:2005bp}. Moreover, a
non-Ermakov-Pinney equation for the same system was also proposed in
form of a non-linear Schr\"{o}dinger-type equation (hereafter-NLS)
and it was found that solutions of the NLS-type equation are
correspondent to solutions of the generalized Ermakov-Pinney
equation \cite{Williams:2005bp,D'Ambroise:2006kg}\footnote{
Considering Bianchi I scalar field cosmology, one can also construct
a corresponding linear Schr\"{o}dinger-type equation by redefining
cosmological quantities \cite{D'Ambroise:2007gm}}. Conclusion of how
to relate NLS quantities to quantities of standard Friedmann
formulation is shown in \cite{Gumjudpai:2007qq} which gives
extension to phantom field case. It also shows that the NLS wave
function is in general non-normalizable. Expressing cosmological
quantities in form of NLS quantities may suggest an alternative way
of solving problems in scalar field cosmology. In such method,
presumed knowledge of scale factor function with time $a(t)$ must be
given first and later one can evaluate NLS potential based on $a(t)$
assumed. Here we attempt to solve for scalar field exact solution
within the NLS framework in various cases with flat and non-flat
spatial geometries. We compare the results to the solutions obtained
in standard Friedmann formulation. Exponential expansion $a \sim
\exp(t/\tau) $ and power-law expansion $a \sim t^q $ are assumed
where $\tau$ are finite characteristic time and $q$ is a positive
constant.

This article is organized as follow. We express our cosmological
system in Sec. \ref{sec:System} before introducing the NLS
formulation in Sec. \ref{sec:NLS}. Afterward, we consider each model
of expansion separately. Exponential expansion is presented in Sec.
\ref{sec:expo} where we obtain exact solution from effective
equation of state and later we solve for exact solution from
Friedmann system. Afterward, in Sec. \ref{expoNS}, we consider NLS
formulation for the exponential expansion and solve for exact
solutions in NLS framework. We analyze effective equation of state
for the exponential expansion case in Sec. \ref{sec:expoanaweff}.
When considering power-law expansion, we work and organize the
contents in the same spirit and order as in the previous sections.
Beginning from Sec. \ref{sec:power}, \ref{sec:powerNLS} and
\ref{sec:poweranaweff}. Finally, conclusion is made in Sec.
\ref{sec:con}.

\section{Cosmological System} \label{sec:System}
To be realistic, two perfect fluids are considered in our system:
barotropic fluid and scalar field fluid. The perfect barotropic
fluid pressure $p_{\gamma}$ and density $\rho_{\gamma}$ obey an
equation of state, $p_{\gamma} = (\gamma-1)\rho_{\gamma} =
w_{\gamma}\rho_{\gamma}$ while for scalar field, $p_{\phi} =
w_{\phi}\rho_{\phi}$. Total density and total pressure are
$\rho_{\rm tot} = \rho_{\gamma} + \rho_{\phi}$ and $p_{\rm tot} =
p_{\gamma} + p_{\phi}$. The effective equation of state is
weighed-value of these two
components, %
\be w_{\rm eff} = \frac{{\rho_{\phi}w_{\phi} + \rho_{\gamma}
w_{\gamma}}}{{\rho_{\rm tot}}}. \label{weff}
\ee %
For the barotropic fluid, its equation of state coefficient
$w_{\gamma}$ is written in term of $n$. We set $w_{\gamma} \equiv
(n-3)/3$ so that $n = 3(1+w_{\gamma}) = 3\gamma$, hence
$w_{\gamma}=-1$ corresponds to $n=0$, $w_{\gamma}=-1/3$ to $n=2$,
$w_{\gamma}=0$ to $n=3$, $w_{\gamma}=1/3$ to $n=4$, and
$w_{\gamma}=1$ to $n=6$. The conservation equation is therefore
\bea \dot{\rho}_{\gamma} = - n H \rho_{\gamma} \label{1} \eea
with solution obtained directly,
\be \rho_{\gamma} = \frac{D}{a^{n}}\,,\label{barorho} \ee therefore
$ p_{\gamma} = [{(n-3)}/{3}]({D}/{a^{n}})\,, $ where $a$ is scale
factor, the dot denotes time derivative, $H = \dot{a}/a$ is Hubble
parameter and $D\geq 0$ is a proportional constant. The scalar field
considered here is minimally coupling to gravity with Lagrangian
density, $ \mathcal{L} = (1/2)\epsilon \dot{\phi}^2 - V(\phi)\,, $
where $\epsilon=1$ for non-phantom case and $-1$ for phantom case.
Density and pressure of the field are given as
\bea \rho_{\phi} = \frac{1}{2} \epsilon \dot{\phi}^2 +
V(\phi)\,,\;\;\;\;\;\;\;\;\; p_{\phi}= \frac{1}{2} \epsilon
\dot{\phi}^2 - V(\phi)\,, \label{phanp}\eea therefore%
\be 
w_{\phi} = \f{p_{\phi}}{\rho_{\phi}} = \frac{\epsilon \dot{\phi}^2
-2V(\phi)}{\epsilon \dot{\phi}^2 + 2V(\phi)}\,.\label{wphi} %
\ee %
The field obeys conservation equation
\be \epsilon\left[\ddot{\phi} + 3H\dot{\phi} \right] + \frac{{\rm
d}V}{{\rm d}\phi} = 0\,. \label{phanflu} \ee

 Considering Friedmann-Lema\^{i}tre-Robertson-Walker (FLRW) universe, the
Friedmann equation and acceleration equation are
\bea H^2 &=& \frac{\kappa^2}{3}\rho_{\rm tot} - \frac{k}{a^2}\,,
\label{fr}
\\ \frac{\ddot{a}}{a} &=&  -\frac{\kappa^2}{6} \rho_{\rm tot}
(1 + 3w_{\rm eff})   \label{ac}\eea 
where $\kappa^2 \equiv 8\pi G = 1/M_{\rm P}^2$, $G$ is Newton's
gravitational constant, $M_{\rm P}$ is reduced Planck mass, $k$ is
spatial curvature. Using Eqs. (\ref{barorho}), (\ref{phanp}),
(\ref{phanflu}) and (\ref{fr}), it is straightforward to show that
\bea   \epsilon \dot{\phi}(t)^2 & = & -\frac{2}{ \kappa^2} \left[ \dot{H} - \frac{k}{a^2}  \right] - \frac{n D}{3  a^n} \,, \label{phigr} \\
V(\phi) &=& \frac{3}{\kappa^2} \left[H^2 + \frac{\dot{H}}{3} +
\frac{2k}{3 a^2} \right] + \left(\frac{n-6}{6}\right)
\frac{D}{a^n}\,. \label{Vgr} \eea %

\section{Non-linear Schr\"{o}dinger-type formulation}
\label{sec:NLS}
Correspondence between non-linear Schr\"{o}dinger formulation for
canonical scalar field cosmology with barotropic fluid was shown in
\cite{D'Ambroise:2006kg} and was concluded recently in
\cite{Gumjudpai:2007qq}. In the Schr\"{o}dinger formulation, wave
function $u(x)$ is related to scale factor in cosmology as
 \bea
 u(x) &\equiv& a(t)^{-n/2}\,, \label{utoa} \eea
 and Schr\"{o}dinger total energy $E$ and Schr\"{o}dinger potential
 $P(x)$ are linked to cosmology as
 \bea
 E &\equiv&  -\frac{\kappa^2 n^2}{12} D \,, \label{E} \\
P(x) &\equiv& \frac{\kappa^2 n}{4}a(t)^{n} \epsilon \dot{\phi}(t)^2
\,. \label{schropotential} \eea
These quantities satisfy non-linear Schr\"{o}dinger-type equation:
 \bea \frac{{\rm d}^2 }{{\rm d}x^2}u(x) + \left[E-P(x)\right]
u(x)
 = -\frac{nk}{2}u(x)^{(4-n)/n}\,.   \label{schroeq} \eea
The mapping from $t$ to $x$ is via $ x = \sigma(t) $,  such that
\bea \dot{x}(t)&=& u(x)\,, \label{dsigtou} \\
\phi(t) &=& \psi(x)\,. \eea
We comment that the relation $\psi'(x)^2 = (4/n \kappa^2)P(x)$ in
Ref. \cite{D'Ambroise:2006kg} does not include phantom field case.
Modification is made in recent work \cite{Gumjudpai:2007qq} so that
the solution includes the phantom field case, therefore 
\bea \psi(x) = \pm
\frac{2}{\kappa\sqrt{n}}\int{\sqrt{\frac{P(x)}{\epsilon}}}\,{\rm
d}x\,  \,.
\label{phitoPx} %
\eea
If $P(x) \neq 0$ and $n \neq 0$. There exists an inverse function of
$\psi(x)$ as $\psi^{-1}(x)$. Therefore $ x(t) = \psi^{-1}\circ
\phi(t) $ and the scalar field potential, $V\circ \sigma^{-1}(x)$
can be expressed as function of time, \be V(t) = \frac{12}{\kappa^2
n^2}\left( \frac{{\rm d} u}{{\rm d}x} \right)^2 - \frac{2
u^2}{\kappa^2 n} P(x) + \frac{12 u^2}{\kappa^2 n^2}E + \frac{3 k
u^{4/n}}{\kappa^2} \,. \label{vt} \ee

\section{Exponential expansion} \label{sec:expo}
\subsection{Solution solved from effective equation of state for $k=0$
case} \label{sec:expophiweff} Exponential expansion reads
 \bea a(t) = \exp{(t/\tau)}\,, \label{expex} \eea where
$\tau$ is a positive constant.  Flat universe undergoes exponential
expansion only when $w_{\rm eff} = -1$. The effective equation of
state, (Eq. (\ref{weff})) with Eqs. (\ref{phanp}) and (\ref{wphi})
can therefore be
written as\footnote{We are not considering a cosmological constant but a dynamical scalar field and a barotropic fluid which together yield $w_{\rm eff} =-1$.} %
\bea %
\epsilon \dot{\phi}^2 &=& - \frac{n}{3}\rho_{\gamma} \,,%
\eea which can be integrated directly, using Eq. (\ref{barorho}), to
\bea \phi(t) = \pm 2\tau\sqrt{\frac{D}{3n}}\,e^{-nt/2\tau} +
\phi_0\,. \label{expophisol} \eea The solution above is obtained
when assuming phantom scalar field, i.e. $\epsilon=-1$. If the
scalar field is not phantom, the solution is imaginary.
\subsection{Solution solved from Friedmann formulation}
\label{sec:expophiexact} 
Another way to find the exact solution is to use Eq. (\ref{expex}),
in Eq. (\ref{phigr}). Therefore
\be \epsilon\dot{\phi}(t)^2 =  \frac{2k}{\kappa^2}\,e^{-2t/\tau} -
\frac{n D}{3}\,e^{-nt/\tau} \,, \label{phigrexpex}\ee which gives an
integration:
\bea \phi(t) = \pm \int\sqrt{\frac{1}{\epsilon}\left(
\frac{2k}{\kappa^2} e^{-2t/\tau} - \frac{nD}{3}e^{-nt/\tau}\right)}
{\rm d}t \,. \label{g1} \eea %

\subsubsection{Simplest case}
In the case of $k=0$ and $D=0$, the integration yields a constant
$\phi_{0}$. Eq. (\ref{weff}) becomes $ w_{\phi} = -1$. This is a
cosmological constant as seen in simplest model of exponential
expansion. When assuming only $k=0$ and $\epsilon = -1$ but with $D
\neq 0$, the solution of Eq. (\ref{g1}) is the same as the Eq.
(\ref{expophisol}) previously. For a scalar field domination in a
non-flat universe ($D=0, k \neq 0$), the solution is
\bea \phi(t) = \pm
\frac{\tau}{\kappa}\sqrt{\frac{2k}{\epsilon}}\,e^{-t/\tau} +
\phi_0\,. \eea where $k$ and $\epsilon$ must have the same sign,
otherwise the solution is imaginary.

\subsubsection{The case of non-zero $k$ and non-zero $D$}
When $k$ and $D$ are both not negligible. Performing integration to
the Eq. (\ref{g1}) is more complicated and could be impossible
unless assumption of barotropic fluid type. When assuming a
particular type of fluid in the integration, i.e. $n = 0, 2, 3, 4$
and $6$, analytical solution can be found for all $n$ vales in
complicated forms. For example, the simplest among these is dust
case ($n=3$) which has solution:
\bea \phi(t) = \pm \f{2 \tau}{3 D
\sqrt{\epsilon}} \left( \frac{2k}{\kappa^2}  - D
e^{-t/\tau}\right)^{3/2}
 + \,\phi_0\,,
\label{expophisolFMdust} \eea
with additional rule that $k \geq 0$
and $\epsilon =1$ otherwise it is imaginary. In the next section, we
will show how to obtain solution in NLS formulation for dust and
radiation cases.

\section{Exponential expansion: Solutions solved with NLS formulation}\label{expoNS}
For exponential expansion, following Eqs. (\ref{utoa}) and
(\ref{dsigtou}), we get
\bea u(x) = \dot{x}(t) = \exp{\left(-{nt}/{2\tau}\right)}\,. \eea
Integrating the above equation, hence parameters $x$ and $t$ scale
as
\bea x(t) =  -\frac{2\tau}{n} e^{-{nt}/{2\tau}} + x_0\,,
\label{Expoxt}\eea
where $x_0$ is an integration constant. The reverse is
\bea t(x) =  -\frac{2\tau}{n} \ln \left[
\left({-{n}/{2\tau}}\right)\left(x - x_0 \right) \right]\,, \eea
where the condition $x < x_0 $ must be imposed. Now we can write
wave function as
\bea u(x) = -\frac{n}{2\tau}(x-x_0)\,, \eea
which is a linear function.  Using Eq. (\ref{phigrexpex}), hence the
Eq. (\ref{schropotential}) reads
\bea%
P(t) & =& \frac{k n}{2}\,e^{(n-2)t/\tau}   - \frac{\kappa^2 n^2
D}{12}\,.  \eea
Here the Schr\"{o}dinger kinetic energy term is \bea T(t) &=&
-\frac{k n}{2}\,e^{(n-2)t/\tau}\,. \eea
Expressing in Schr\"{o}dinger formulation, these functions are
written in term of $x$,
\bea P(x)   & = & \frac{kn}{2}\left[ -\frac{n}{2\tau}(x-x_0)
\right]^{-2(n-2)/n}  - \frac{\kappa^2 n^2 D}{12}\,, \label{Px} \\
 T(x) &= & -\frac{kn}{2}\left[ -\frac{n}{2\tau}(x-x_0)
\right]^{-2(n-2)/n}\,. \eea
In order to obtain the scalar field potential $V(t)$, we use Eqs.
(\ref{utoa}), (\ref{E}), (\ref{schropotential})  in Eq. (\ref{vt}),
we finally obtain
\be V(t) = \frac{3}{\kappa^2 \tau^2} + \frac{2k}{\kappa^2}\,
e^{-2t/\tau} + \left(\frac{n-6}{6}\right) D\,e^{-nt/\tau}\,.
\label{vtexpo}\ee which is checked by using Eq. (\ref{expex}) in
standard formula (\ref{Vgr}).  We use Eq. (\ref{Px}) in Eq. (\ref{phitoPx}), then %
\bea %
\psi(x)  =  \frac{\pm 2}{\kappa \sqrt{n}} \times  \int
{\sqrt{\frac{kn}{2\epsilon}\left[ \frac{-n}{2\tau}(x-x_0)
\right]^{-2(n-2)/n} - \frac{\kappa^2n^2 D}{12 \epsilon}}}\:
{\rm d}x  \,. \label{q2} %
\eea %
We will integrate this equation in cases of $k=0$ and $k\neq 0$. 
\subsection{The case $k=0$} When $k=0$ and $D\n 0$ integrating Eq. (\ref{q2})
and transforming $x$ to $t$ with Eq. (\ref{Expoxt}) yields same
result as Eq. (\ref{expophisol}) obtained by solving effective
equation of state equation or by integrating from the Friedmann
formulation. Real solution exists only when the scalar field is
phantom. With the solution (\ref{expophisol}), the scalar field
potential in term of $\phi$, reads
\bea V(\phi)& =& \frac{3}{\kappa^2 \tau^2} +
\left(\frac{n-6}{6}\right)\frac{3n}{4\tau^2}(\phi-\phi_0)^2\,. \eea
\subsection{The case $k\neq 0$} When $k\neq 0 $ and $D\neq 0$, the integral
(\ref{q2}) can be integrated yielding complicated hypergeometric
function even when $n$ is not specified.  The case $n=0$ is excluded
from our consideration by the reason mentioned in Sec \ref{sec:NLS}.
For naturalness, we consider radiation ($n=4$) and dust ($n=3$).
\subsubsection{Radiation case}
Radiation fluid corresponds to $n=4$, the integral (\ref{q2})
becomes
\bea \psi(x) &=&  \pm \frac{1}{\kappa} \int
\sqrt{-\frac{k\tau}{\epsilon}\frac{1}{(x-x_0)} -
\frac{4}{3}\frac{\kappa^2 D}{\epsilon} }\, {\rm d}x \,. \eea
Here $x$ could be negative, $\epsilon$ can possibly be either
$\pm1$. The solution in radiation case is
\bea \psi(x) &=& \pm
\sqrt{\frac{1}{\epsilon}\left[ -\frac{4}{3} D (x-x_0)^2 - \frac{k
\tau}{\kappa^2} (x-x_0)
    \right] } \nonumber  \\
   & & \pm \frac{k\tau}{4\kappa^2} \sqrt{\frac{3}{
    D\epsilon}} \arctan\left\{{\frac{[8\kappa^2 D (x-x_0)/3\epsilon] +
    k\tau/\epsilon}{ [4\kappa \sqrt{D}/(\epsilon\sqrt{3})]\sqrt{-[4\kappa^2 D(x-x_0)^2/3] - k\tau(x-x_0)}
    }}\right\} + \psi_0\,, \nonumber \\
\eea 
allowing only $\epsilon = 1$ case for the solution to be real.
Transforming $x$ scale to the $t$ scale using Eq. (\ref{Expoxt}),
the solution therefore reads%
\bea \phi(t) &=& \pm\sqrt{\frac{1}{\epsilon}  \left(
-\frac{D\tau^2}{3} \,e^{-4t/\tau} + \frac{k \tau^2}{2\kappa^2}\,
e^{-2t/\tau} \right)} \nonumber \\ & &
\pm\frac{k\tau}{4\kappa^2}\sqrt{\frac{3}{D \epsilon}}  \times
\arctan{\left\{   \frac{-[4\kappa^2 D
\tau/(3\epsilon)]\,e^{-2t/\tau} + k \tau/\epsilon}{ [4\kappa
\sqrt{D}/(\epsilon\sqrt{3})] \sqrt{-(\kappa^2 D
\tau^2/3)\,e^{-4t/\tau} + (k\tau^2/2)\,e^{-2t/\tau}   } }
 \right\}} + \phi_0 \,.
\label{expophisolNLSrad}%
\eea
The solution above, when assuming $k=0$, reduces to the solution
(\ref{expophisol}) when $n=4$, confirming the correctness of the
result obtained.

\subsubsection{Dust case}
The integral (\ref{q2}) in the dust case $n=3$ reads
\bea \psi(x) = \pm\frac{2}{\kappa\sqrt{3}} \int
\sqrt{\left(\frac{3}{2}\tau^2\right)^{1/3} \frac{k}{\epsilon}
\frac{1}{(x-x_0)^{2/3}}-\frac{3}{4} \frac{\kappa^2
D}{\epsilon}}\,{\rm d}x \,,  \eea 
with solutions
\bea \psi(x) = \pm \s{\f{D}{\e}} \left[
\left(\frac{3\tau^2}{2}\right)^{1/3} \frac{4 k}{3 \kappa^2 D} -
(x-x_0)^{2/3} \right]^{3/2} + \,\psi_0\,. \eea
With similar procedure to the radiation case, using (\ref{Expoxt}), the solution is therefore,%
\bea
\phi(t) = \pm \f{2 \tau}{3 D \sqrt{\epsilon}} \left(
\frac{2k}{\kappa^2}  - D e^{-t/\tau}\right)^{3/2}
 + \,\phi_0\,, \label{expophisolNLSdust}
\eea %
which is the same as Eq. (\ref{expophisolFMdust}) derived from
Friedmann formulation. This solution when assuming $k=0$ is exactly
the same as the solution (\ref{expophisol}) when $n=3$ (dust fluid).
This also confirms that our results from NLS formulation are
correct. The NLS solution can solve the case when $k$ and $D$ are
non-zero together without knowing $n$ value while the standard
procedures in Sec.  \ref{sec:expophiexact} can not unless assuming a
particular value  $n = 0, 2, 3, 4, 6$. However, it must be noticed
that one can not reduce the NLS solutions (\ref{expophisolNLSrad})
and (\ref{expophisolNLSdust}) to the $D=0$ case directly since there
are mixed multiplication term of $n$ and $k$ in the solution and
also the value of $n$ has already been put in. Hence setting $D=0$
in (\ref{expophisolNLSrad}) and (\ref{expophisolNLSdust}) can not be
considered as a pure scalar field dominant case.

\section{Exponential expansion: Analysis on effective equation of
state coefficient} \label{sec:expoanaweff}
 The exponential expansion in our scenario is caused from mixed effect of
 fluids and spatial curvature. We discuss mixed effect on equation of
state here. Definition of effective equation of state coefficient, $
w_{\rm eff} = ({\rho_{\phi}w_{\phi} + \rho_{\gamma}
w_{\gamma}})/{\rho_{\rm tot}}\,$ together with Eqs. (\ref{phanp}),
(\ref{phigrexpex})  and (\ref{vtexpo}) in context of exponential
expansion becomes
 \be w_{\rm eff} = \frac{-1 - (k\tau^2/3)e^{-2t/\tau}}{1+k\tau^2 e^{-2t/\tau}}\,, \label{Expoweffkt}\ee
 which is infinite when
 \be t= \f{\tau}{2}\ln(-k\tau^2)\,. \ee
Infinity can possibly happen only when $k=-1$ because logarithm
function forbids negative domain. In order to acquire exponential
expansion in flat universe, one needs to have $w_{\rm eff} = -1$,
but this is not true when $k$ term is non-trivial.
Therefore we can only express $w_{\phi}$ in term of $w_{\rm eff}$ as %
\be %
w_{\phi} = \frac{[\,(3k/\kappa^2) e^{-2t/\tau} + 3/(\kappa^2
\tau^2)\,] w_{\rm eff} \:-\: [(n-3)/3]\, D
e^{-nt/\tau}}{(3k/\kappa^2) e^{-2t/\tau} \,+ \,3/(\kappa^2 \tau^2)\,
-\, D e^{-nt/\tau}}\,, \label{Expophikt}\ee %
for exponential expansion. The Eq. (\ref{Expoweffkt}) does not
depend on properties ($n$) or amount ($D$) of the barotropic fluid.
It reduces to $w_{\rm eff} = -1$ when $k=0$ as expected. Considering
Eq. (\ref{Expophikt}), if $D=0$ and $k=0$, it yields $w_{\phi} =
w_{\rm eff}$ while setting $D=0$ alone also gives the same result.

\section{Power-law expansion} \label{sec:power}

\subsection{Bound value of $\phi(t)$ from effective equation of state for $k=0$
case} \label{sec:powerphiweff} 
In power-law expanding universe, scale factor evolves with time as
 \bea a(t) = t^q \,. \label{Powerat} \eea
where $q>0$ is a constant.  In flat ($k=0$) universe, it is known
that the power-law expansion, is attained when $-1< w_{\rm eff} <
-1/3$ where $q = 2/[3(1+w_{\rm eff})]$. The effective equation of
state, (Eq. (\ref{weff})) hence is a condition
\bea %
-\f{n}{3}\rho_{\gamma}\; < \; \epsilon \dot{\phi}^2  \; < \;
 \frac{2}{3}\left( \f{\epsilon \dot{\phi}^2}{2} +V  \right)    + \left(\f{2-n}{3}\right)\rho_{\gamma} \,,%
\eea %
i.e. $0 \:<\: {\epsilon \dot{\phi}^2} + \rho_{\gamma}n/3 \: < \:
(2/3)\rho_{\rm tot}$. Both values of $\epsilon$ can be assigned and
the power-law expansion is sustained as long as the condition is
satisfied.

\subsection{Solution solved from Friedmann formulation}
\label{sec:powerphiFr} 
If we directly consider Eq. (\ref{phigr}), the solution for
power-law expansion is an integration:
\bea \phi(t) = \pm \int\sqrt{\frac{1}{\epsilon}\left(
\f{2q}{\kappa^2 t^2} +
  \frac{2k}{\kappa^2 t^{2q}} - \frac{nD}{3 t^{qn}}\right)} {\rm d}t
\,. \label{powerintfri} \eea

\subsubsection{Simplest case}
Simplest integration case is when $k=0$ and $D=0$. The solution of
Eq. (\ref{powerintfri}) is well known \cite{Lucchin}, %
\bea \phi(t) = \pm \sqrt{\f{2q}{\epsilon\kappa^2}} \ln t + \phi_{0}
 \,,  \label{simplestPhi} \eea provided that $q$ and $\epsilon$ have the same sign.
Considering power-law inflation, the WMAP five-year combined
analysis based on flat and scalar field domination assumption yields
$q>60$ at more than 99 \% of confident level otherwise excluded
while $q \sim 120$ is at boundary of 68\% confident level
\cite{Komatsu:2008hk}. These results base on single field model
which we can applied the above solution to.
 When
assuming only $k=0$ with $D \neq 0$, the solution of Eq.
(\ref{powerintfri}) is 
\bea %
\phi(t)  =  \pm \f{1}{qn-2} \sqrt{\f{2q}{\epsilon \kappa^2}}\left\{
\ln \left[
 \f{\,t^{-qn+2}}{\left(1 + \sqrt{1-(nD\kappa^2/6q)\,t^{-qn+2}}\right)^2}  \right]
+ 2\sqrt{1-\left(\f{nD\kappa^2}{6q}\right)\,t^{-qn+2}}  +
\ln\left(-\f{nD\kappa^2}{6q}\right) \right\} \:+\:\phi_{0}\,, \label{PowerK0FriSol}\nonumber \\
\eea
where, when $q = 2/n$, the field has infinite value. The last
logarithmic term in the bracket is an integrating constant which is
valid only when $q<0$. To attain power-law expansion, $q$ must be
positive. Hence, this term is not defined for power-law expansion.
We will see later that the NLS result does not have this problem.
For the reverse case, $D=0, k \neq 0$, the solution is
\bea \phi(t) = \pm \f{1}{q-1} \sqrt{\f{2q}{\epsilon \kappa^2}}
  \left\{ \ln \left[
 \f{t^{q-1}}{\sqrt{k/q}}\left(1+ \sqrt{\left(\f{k}{q}\right)t^{-2q+2} + 1} \: \right)  \right]
 - \sqrt{\left(\f{k}{q}\right)t^{-2q+2} + 1}\:  \right\} +
\phi_0\,,
\eea %
which becomes infinite when $q=1$. The values of $q, k$  and
$\epsilon$ must have the same sign in all terms of the solution
otherwise becoming imaginary. Hence, for $q>0$, the condition for
the solution to be valid is $k=1$ and $\epsilon=1$.

\subsubsection{The case of non-zero $k$ and non-zero $D$}
When considering non negligible value of both $k$ and $D$, the Eq.
(\ref{powerintfri}) can not be integrated analytically except when
setting $n=2\;\, (w_{\gamma}=-1/3)$ which is not natural fluid.
Hence it is not considered.

\section{Power-law expansion: Solutions solved with NLS formulation}\label{sec:powerNLS}
Power-law expansion cosmology in NLS-type formulation is presented
and concluded in \cite{Gumjudpai:2007qq}. Important functions needed
for evaluating the field exact solutions are
 \bea
 x &=& \sigma(t) =
-\frac{t^{-\beta }}{\beta} + x_0\,, \label{PWx}
\\
t(x) &=& \frac{1}{\left[-\beta (x-x_0) \right]^{1/\beta}}
\label{PWt}\,,
\\
 \epsilon\dot{\phi}(t)^2 &=& \frac{2q}{\kappa^2 t^2}  +
\frac{2k}{\kappa^2 t^{2q}} - \frac{n D}{3 t^{qn}}\,,
\label{phigrpowerex}
\\
P(x) &=& \frac{2qn}{(qn-2)^2}\frac{1}{(x-x_0)^2} \,+\,
\frac{kn}{2}\left[ \frac{-2}{(qn-2)(x-x_0)} \right]^{2q(n-2)/(qn-2)}
\,- \, \frac{\kappa^2 n^2 D }{12}\,, \label{PWP}
\\
 V(t) &=& \frac{q(3q-1)}{\kappa^2 t^2} + \frac{2k}{\kappa^2 t^{2q}} +
\left(\frac{n-6}{6}\right)\frac{D}{t^{qn}}\,. \label{vtpowerlaw}
\eea
We use Eq. (\ref{PWP}) in Eq. (\ref{phitoPx}), then %
\bea \psi(x)  =  \frac{\pm 2}{\kappa \sqrt{n}} \times  \int {\sqrt{
\f{2qn}{\e(qn-2)^2}\f{1}{(x-x_0)^2} + \f{kn}{2\e} \left[
\f{-2}{(qn-2)}\f{1}{(x-x_0)} \right]^{2q(n-2)/(qn-2)} -
\frac{\kappa^2n^2 D}{12 \epsilon}}}\: {\rm d}x  \,. \label{PowerPsi}
\eea %
We consider the solution in cases of $k=0$ and $k\neq 0$. Recall
that setting $D= 0$ can not be considered as an absence of
barotropic fluid due to existence of $n$ in the other terms.
\subsection{The case $k=0$}
Solution to the integral (\ref{PowerPsi}) for $k=0$ case is
\bea
\psi(x) &=& \pm  \s{\f{8q}{\e \kappa^2 (qn-2)^2}} \times
\nonumber
\\ & &\left\{ - \s{1 - \left[  \f{\kappa^2 D n(qn-2)^2}{24q}
(x-x_0)^2 \right]} + \ln \left[ \f{1+ {\s{1 - \left[ {\kappa^2 D n
(qn-2)^2}/{24 q}\right] (x-x_0)^2 }}}{ (x-x_0)} \f{4qn}{\e (qn-2)^2}
\right] \right\}\,. \nonumber \\ \eea Transforming to the $t$
variable using Eq. (\ref{PWx}), we obtain,
\bea \phi(t) & = & \pm
\f{1}{qn-2} \sqrt{\f{2q}{\epsilon \kappa^2}}\left\{ \ln \left[
 \f{\,t^{-qn+2}}{\left(1 + \sqrt{1-(nD\kappa^2/6q)\,t^{-qn+2}}\right)^2}  \right]
+ 2\sqrt{1-\left(\f{nD\kappa^2}{6q}\right)\,t^{-qn+2}}  +
\ln\left(\f{qn-2}{2qn}\right)^2 \right\} \:+\:\phi_{0}\,. \label{PowerK0NLSSol} \nonumber \\
\eea
This solution differs from the solution (\ref{PowerK0FriSol}) only
the last logarithmic term in the bracket which is only an
integrating constant. When $q = 2/n$ or $n =0$, the field has
infinite value. The last logarithmic term does not restrict the sign
of $q$. Only $q$ and $\epsilon$ must have the same sign for the
solution to be real.

\subsection{The case $k\neq 0$}
In case of non-zero $k$ and non-zero $D$, the integral
(\ref{PowerPsi}) can not be integrated analytically even when
assuming each $n$ value except when $n = 2$ which is not natural
fluid.

\section{Power-law expansion: Analysis on effective equation of
state coefficient}\label{sec:poweranaweff}
 Similar to the analysis in Sec. \ref{sec:expoanaweff}, mixed effect of
 the two fluids and spatial curvature results in power-law
 expansion. The coefficient $ w_{\rm eff}$, with Eqs. (\ref{phanp}), (\ref{phigrpowerex}) and
(\ref{vtpowerlaw}), reads
 \be w_{\rm eff} = \frac{(-3q^2 + 2q)\,t^{2q-2}-k}{3q^2\,t^{2q-2}+3k}\,, \label{Powerweffkt}\ee
 which becomes infinity if
 \be t= \left(\f{-k}{q^2}\right)^{1/(2q-2)}\,. \ee
We can also express $w_{\phi}$ in term of $w_{\rm eff}$ as %
\be
w_{\phi} = \frac{[ (3q^2/\kappa^2) t^{-2} + (3k/\kappa^2)
t^{-2q}
\,] w_{\rm eff} \:-\: [(n-3)/3]\, D t^{-qn}}{ (3q^2/\kappa^2) t^{-2} + (3k/\kappa^2) t^{-2q} -D t^{-qn}    }\,, \label{Powerwphikt}\ee %
for power-law expansion. The Eq. (\ref{Powerwphikt}), when $D=0$ and
$k=0$, yields $w_{\phi} = w_{\rm eff}$ as expected. Similar to the
case of exponential expansion, setting $D=0$ alone also yields
$w_{\phi} = w_{\rm eff}$.  In flat universe, power-law expansion
happens when $w_{\rm eff} $ lies in an interval  $(-1, -1/3)$. But
in $k \neq 0$ universe, it is no longer true. Considering flat
universe, setting $k=0$ in Eq. (\ref{Powerweffkt}) yields $q =
2/[3(1+w_{\rm eff})]$. The condition $-1< w_{\rm eff} < -1/3$
therefore corresponds to $q>0$ as known. The condition also yields
\bea -1 - (1+w_{\gamma})\f{\rho_{\gamma}}{\rho_{\phi}} \, < \,
w_{\phi} \, < \, -\f{1}{3} - \left(\f{1}{3} +
w_{\gamma}\right)\f{\rho_{\gamma}}{\rho_{\phi}}\,. \eea
If there is
more non-negligible radiation fluid (with $w_{\gamma} = 1/3$), it is
noticed that the interval shifts to the more left. For example,
setting $\rho_{\gamma} = 0.1\rho_{\phi}$, the interval shifts to
about $-1.133 < w_{\phi} < -0.4 $. If we assume more realistic
situation when dust (dark matter and other matter elements) is
presented. The dust density and dark energy is about 28\% and 72\%
of total density, therefore
 $\rho_{\gamma} \simeq (28/72) \rho_{\phi} \simeq 0.389 \rho_{\phi}$, the
interval is $-1.389 < w_{\phi} < -0.463$ which covers valid range of
recent observational data, assuming dynamical $w$ with flat
universe, $-1.38 < w_{\phi,0} < -0.86$ at 95\% confident level
\cite{Komatsu:2008hk}.

\section{Conclusions} \label{sec:con}
This letter reports and demonstrates a method of solving for
canonical scalar field exact solution in a non-flat FLRW universe
with barotropic fluid using non-linear Schr\"{o}dinger (NLS)-type
formulation in comparison with the method in the standard Friedmann
framework. We consider phantom and non-phantom scalar field cases
with exponential and power-law accelerating expansion. We evaluate
all NLS quantities needed to find the solution, e.g.
non-normalizable wave function and Schr\"{o}dinger potential. Our
process is reverse to a problem solving in quantum mechanics that
the wave function is expressed first by the expansion function,
$a(t)$ before evaluating the Schr\"{o}dinger potential based on a
known expansion function.  In NLS formulation the total energy $E$
is negative. We do an analysis on effective equation of state to
both cases of expansion. We expresses $w_{\rm eff}$ in term of $q$
and $k$. In a flat universe, in order to have power-law expansion,
the interval $(-1, -1/3)$ of the $w_{\rm \phi}$, is shifted leftward
to more negative if more barotropic fluid density is presented.

Within framework of the standard Friedmann formulation, we obtained
exact solution in various cases. Later we solved the problem using
NLS formulation, in which the wave function is equivalent to the
scalar field exact solution. NLS method is restricted by the fact
that its scalar field solution is valid only when the barotropic
fluid density is presented. Setting $D=0$ does not imply the absence
of barotropic fluid because the barotropic fluid parameter $n$ still
appears in the other terms of the Schr\"{o}dinger potential.
Therefore NLS formulation can not be applied to situation when the
scalar field is dominant and $D\sim 0$. Hence it is more suitable
for a system of dark energy and dust dark matter fluid. This is a
disadvantage point of the NLS formulation. Transforming from
standard Friedmann formulation to NLS formulation makes
 $n$ appear in all terms of the integrand and also changes fluid
 density
 term $D$ from time-dependent term to a constant $E$.
Hence the number of $x$(or equivalently $t$)-dependent terms is
reduced by one. This is a good aspect of the NLS. In both Friedmann
formulation and NLS formulation, the solutions when $k\neq 0 $ and
$D\neq 0$ are difficult or might be impossible to solve unless
assuming values of $q$ and $n$. Hence reduction number of
$x$-dependent term helps simplifying the integration. There are also
other good aspects of NLS formulation. Firstly, in the case of
exponential expansion with NLS formulation, the solution when $k\neq
0$ and $D\neq 0$ can be obtained without assuming $n$ value while
$n=0, 2, 3, 4, 6$ must be given if working within Friedmann
formulation. Secondly, for power-law expansion with $k = 0$, the
result (\ref{PowerK0NLSSol}) obtained from NLS formulation has
integrating constant that does not restrict $q$ value while
(\ref{PowerK0FriSol}) obtained from Friedmann formulation needs
$q<0$ which violates power-law expansion condition ($q>0$). For
power-law expansion, the most difficult case is when $k\neq 0$ with
$D\neq 0$. In both formulations, the integral can not be integrated
unless assuming $n=2$ (equivalent to $w_{\gamma} = -1/3$) which is
not a physical fluid. We introduce here alternative method to obtain
scalar field exact solution with advantage over and disadvantage to
standard Friedmann formulation. The NLS formulation could render
more interesting techniques for scalar field cosmology.

\begin{acknowledgements}
B.~G. is a TRF Research Scholar of the Thailand Research Fund. B.~G.
gives a special thank for hospitalities, supports during his
research visits and for seminal invitations, to Yupeng Yan, Prapan
Manyum and Chinorat Kobdaj at Suranaree University of Technology, to
Department of Physics of Ubon Rajathanee University and to the
organizers of the Forth Aegean Summer School on Black Holes \& the
First Annual School of the EU Network {\it UniverseNet}-The Origin
of the Universe in Mytilini, Island of Lesvos, Greece where this
work was partially completed.
\end{acknowledgements}

\end{document}